\documentclass[aps,preprint]{revtex4}

\usepackage{graphicx}
\draft

\begin{document}

\draft

\preprint{\today}
\title{Structure and transport properties of amorphous aluminium silicates: 
       computer simulation studies}
\author{Patrick Pfleiderer, J\"urgen Horbach, and Kurt Binder}
\affiliation{Institut f\"ur Physik, Johannes--Gutenberg--Universit\"at Mainz,
Staudinger Weg 7, D--55099 Mainz, Germany}

\begin{abstract}
The structure and transport properties of SiO$_2$--Al$_2$O$_3$ melts
containing 13\,mol\% and 47\,mol\% Al$_2$O$_3$ are investigated by
means of large scale molecular dynamics computer simulations. The
interactions between the atoms are modelled by a pair potential which
is a modified version of the one proposed by Kramer {\it et al.}
[J.~Am.~Chem.~Soc.~{\bf 64}, 6435 (1991)]. Fully equilibrated melts
in the temperature range 6000\,K\,$\ge T > 2000$\,K are considered as
well as glass configurations, that were obtained by a rapid quench from
the lowest melt temperatures. Each system is simulated at two different
densities in order to study the effect of pressure on structural and
dynamic properties. We find that the Al atoms are, like the Si atoms,
mainly four--fold coordinated by oxygen. However, the packing of the
AlO$_4$ tetrahedra is very different from that of the SiO$_4$ tetrahedra,
which is reflected by the presence of triclusters (O atoms surrounded by
three cations) and edge--sharing AlO$_4$ tetrahedra. On larger length
scales, a micro--segregation occurs, resulting in an Al--rich network
percolating through the Si--O network. This is reflected in a prepeak
of concentration--concentration structure factors around 0.5\,\AA$^{-1}$
(both in the system with 47\,mol\% and 13\,mol\% Al$_2$O$_3$!).  We also
address the interplay between structure and mass transport.  To this end,
the behavior of the selfdiffusion constants for the different compositions
and densities is studied.
\end{abstract}

\maketitle
%\pacs{PACS numbers: 61.20.Lc, 61.20.Ja, 02.70.Ns, 64.70.Pf}

%
%
\section{Introduction}
\label{sec1}

One of the most abundant oxides in natural silicates and technological
silicate glasses is Al$_2$O$_3$. Although alumino silicates are
therefore of central interest in geosciences and materials science,
the structure even of the binary system SiO$_2$--Al$_2$O$_3$ is far from
being well--understood.  However, the knowledge of the chemical ordering
in the latter system provides also the basis for a better understanding
of the structure of more geologically relevant alkali or alkaline earth
alumino--silicates.

The chemical ordering of aluminium is very complicated
when built into the tetrahedral Si--O network. In order to achieve
local charge neutrality, Al$^{3+}$ ions need a different environment
of O$^{2-}$ ions than Si$^{4+}$ ions.  Thus, unlike SiO$_2$,
Al$^{3+}$ ions do not form a network of AlO$_4$ tetrahedra that are
connected with each other via the O atoms at the corners. Instead,
different experimental techniques such as nuclear magnetic resonance
(NMR) [Lee and Stebbins, 2000; Schm\"ucker {\it et al.}, 1999; Sen and
Youngman, 2004; Stebbins {\it et al.}, 2005; Stebbins and Xu, 1997;
Xue and Kanzaki, 1999] as well as IR and Raman spectroscopy and X--ray
scattering [Morikawa {\it et al.}, 1982; Okuno {\it et al.}, 2005] found
evidence for structural units such as three--fold coordinated oxygen atoms
and five-- and six--fold coordinated aluminium atoms that are not found in pure
silica, unless one considers amorphous silica at very high temperatures,
say above 4000\,K (see Horbach and Kob, 1999).  

At low Al$_2$O$_3$ concentrations, the Al atoms are mainly four--fold
coordinated by oxygens, but, as proposed in NMR studies, these AlO$_4$
units are accompanied by so--called triclusters, i.e.~structural units
where an oxygen atom is surrounded by three cations (where at least
one of them is an Al atom).  Recently, molecular orbital calculations
confirmed the possibility of such triclusters (Kubicki and Toplis, 2002).
Moreover, a combination of molecular dynamics computer simulations and
Hartree--Fock calculations (Tossell and Cohen, 2001; Winkler {\it et
al.}, 2004; Tossell and Horbach, 2005) found evidence that O tricluster
atoms participate typically in two--fold rings (i.e.~edge--sharing
geometries) of composition Al$_2$O$_2$ or AlSiO$_2$.  Certainly, further
experimental studies are necessary to clarify to what extent triclusters
and edge--sharing tetrahedra exist in real aluminium silicates.

As shown by recent NMR studies, five-- and six--fold coordinated
Al atoms are also important structural units in the system
SiO$_2$--Al$_2$O$_3$. The fraction of these highly coordinated Al atoms
tends to increase with increasing Al$_2$O$_3$ concentration (Sen and
Youngman, 2004; Stebbins {\it et al.}, 2005).  All these experimental
findings show that, in alumino silicates, the local chemical ordering
of Al atoms is very different from that of Si atoms.

In a recent MD simulation study of the system (Al$_2$O$_3$)2(SiO$_2$)
[abbreviated in the following as AS2], we have shown that the different
chemical ordering of Al and Si on local length scales also leads
to structural ordering on intermediate length scales of the order of
1\,nm (Winkler {\it et al.}, 2004). This intermediate range order (IRO)
can be described by a microphase separation where the Al--rich network
structure percolates through the Si--O network. The IRO
gives rise to a prepeak in partial static structure factors at a wave
number $q=0.5$\,\AA$^{-1}$.  It can be seen as a precursor of the
metastable liquid--liquid phase separation below $\approx 1900$\,K
that is found experimentally (MacDowell and Beall, 1969) between 
about 10\,mol\% and 50\,mol\% (AS2, with 33\,mol\% Al$_2$O$_3$,
lies approximately in the center of the demixing region).

The formation of IRO seems to be quite a general feature of
multicomponent silicate melts.  Similar IRO as the one found in AS2 has
also been seen in neutron scattering experiments of a calcium silicate
(Gaskell {\it et al.}, 1991), sodium silicates (Meyer {\it et al.}, 2002;
Meyer {\it et al.}, 2004), and alkali aluminosilicates (Cormier {\it et al.},
2001; Kargl and Meyer, 2005).  For the example of sodium silicates, the
IRO is reflected in a prepeak around 0.9\,\AA$^{-1}$ in static structure
factors. It has its origin in the formation of sodium--rich channels in
the static structure. These channels serve as preferential pathways in
an immobile Si--O matrix and thus provide an explanation for the high
mobility of sodium in ion--conducting sodium silicate melts (Horbach {\it et al.},
2002; Meyer {\it et al.}, 2004). The presence of diffusion channels
in sodium silicates is a nice example for the interplay between 
structure and mass transport in glassforming melts.

The structure--transport relation is also a central issue of the
present study.  We have extended our previous simulations of AS2
to aluminium silicates with 13\,mol\% Al$_2$O$_3$ and 47\,mol\%
Al$_2$O$_3$. These systems are simulated at different densities
to see how pressure affects the structural and dynamic properties
of aluminium silicates. Furthermore, we aim to understand
how the interplay between local structural features (triclusters,
two--fold rings, five-- and six--fold coordinated Al ions) and IRO
changes with composition.  This is in turn will allow us to elucidate the interplay
between structural features and transport processes, when we investigate the
temperature dependence of the selfdiffusion constants
for the different systems under consideration.

\section{Model and details of the simulations}
\label{sec2}
As in our recent study of AS2, we use a potential proposed by Kramer
{\it et al.}~(1991) to model the interactions between the atoms. It is
based on the so--called BKS potential (van Beest {\it et al.}, 1990)
for pure silica that has been extensively applied in recent studies
(see, e.g., Horbach and Kob, 1999, and references therein). Apart from
mixtures of SiO$_2$ with Al$_2$O$_3$, the Kramer potential allows also
the consideration of sodium silicates. As shown recently, it provides
quite a realistic description of sodium di--, tri-- and tetra--silicate
(Ispas {\it et al.}, 2002; Meyer {\it et al.}, 2004) as well as of AS2
(as far as comparison with experimental data is possible, see also below)
[Winkler {\it et al.}, 2004; Tossell and Horbach, 2005].

The potential has the following functional form:
\begin{equation}
\phi_{\alpha \beta}(r)=
\frac{q_{\alpha} q_{\beta} e^2}{r} + 
A_{\alpha \beta} \exp\left(-B_{\alpha \beta}r\right) -
\frac{C_{\alpha \beta}}{r^6}
\label{eq1}
\end{equation}
with $\alpha, \beta = {\rm Si, Al, O}$.  Here $r$ is the distance between
an ion of type $\alpha$ and an ion of type $\beta$. 
The values of the parameters
$\{A_{\alpha\beta},B_{\alpha\beta},C_{\alpha\beta}\}$ that were
calculated by {\it ab initio} methods are $A_{\rm SiO}=18003.7572$~eV,
$A_{\rm AlO}=8566.5434$~eV, $A_{\rm OO}=1388.7730$~eV, $B_{\rm
SiO}=4.87318$~\AA$^{-1}$, $B_{\rm AlO}=4.66222$~\AA$^{-1}$, $B_{\rm
OO}=2.76$~\AA$^{-1}$, $C_{\rm SiO}=133.5381$~eV\AA$^{6}$, $C_{\rm
AlO}=73.0913$~eV\AA$^{6}$, and $C_{\rm OO}=175.0$~eV\AA$^{6}$ (for the
Si--Si, Si--Al and Al--Al interactions the latter parameters are all
set to zero) [Kramer {\it et al.}~(1991)].

The Buckingham part of the potential,
\begin{equation}
\phi_{\rm B}(r)=
A_{\alpha \beta} \exp\left(-B_{\alpha \beta}r\right) -
\frac{C_{\alpha \beta}}{r^6} \ ,
\label{eq1b}
\end{equation}
has been truncated and shifted to zero at $r_{\rm c}^{\rm B}=5.5$\,\AA.
In order to make the truncated potential, $\phi_{\rm B}^{\rm trunc}(r)$,
differentiable at $r_{\rm c}^{\rm B}$ it has been multiplied by a
smoothing function of exponential form:
\begin{equation}
\phi_{\rm B}^{\rm trunc}(r) = \left\{
   \begin{array}{l@{\quad \quad}l}
   \left[ \phi_{\rm B}(r)-\phi_{\rm B}(r_{\rm c}^{\rm B}) \right]
   \exp\left( - \frac{d_{\rm B}}{(r-r_{\rm c}^{\rm B})^2} \right) & 
   r < r_{\rm c}^{\rm B} \\
   0 & r \geq  r_{\rm c}^{\rm B} 
 \end{array} \right. \label{eq1c}
\end{equation}
with $d_{\rm B}=0.05$\,\AA$^2$. The parameter $d_{\rm B}$ is chosen such
that the exponential in Eq.~(\ref{eq1c}) does not affect the system's
properties.  But it smoothens out the cusp at $r_{\rm c}^{\rm B}$ which
would lead to a discontinuous force at $r_{\rm c}^{\rm B}$ and thus to a
drift in the total energy in microcanonical MD runs [Allen and Tildesley,
1987]. By the systematic use of smoothing functions as in
Eq.~(\ref{eq1c}) (see also below), we have not encountered an energy drift
in any of the microcanonical runs that we have performed in this work.

In the long--ranged Coulomb--part the charges $q_{\alpha} e$ ($e$:
charge of an electron) are not the bare ionic charges of ions of type
$\alpha$ but are considered to be effective charges.  The charges for
silicon and oxygen are set to $q_{\rm Si}=2.4$ and $q_{\rm O}=-1.2$,
respectively. With the original charge for aluminium, $q_{\rm Al}=1.9$,
the unphysical situation of a non--zero net charge results.  Only with
the additional component phosphorus the system recovers charge neutrality
according to the parameter sets in Kramer {\it et al.} (1991). We have
therefore modified the Kramer potential by using a distance--dependent
Al charge as follows:
\begin{equation}
  q_{\rm Al}(r)= \left\{
    \begin{array}{l@{\quad \quad}l}
       \tilde{q}_{\rm Al} \left[ 1+
       \ln \left( C_{\rm Al} 
       \frac{(r-r_{\rm Al})^2}{1+(r-r_{\rm Al})^2} +1 \right)       
         \right] 
       \exp\left( - \frac{d_{\rm Al}}{(r-r_{\rm c}^{\rm B})^2} \right) 
      & r < r_{{\rm Al}} \\
      \tilde{q}_{\rm Al}  & r \geq r_{{\rm Al}}
    \end{array} \right. \label{eq1d}
\end{equation}
with $\tilde{q}_{\rm Al}=1.8$. The concept of a ``distance--dependent
Al charge'' should not be taken too literally. We just introduce an
additional short--ranged potential to the original one, with a cut--off
at $r=r_{{\rm Al}}$.

The exponential function in Eq.~(\ref{eq1d}) serves again as a smoothing
factor.  The parameters in Eq.~(\ref{eq1d}) are adjusted such that
the potential is very close to the original one at short distances.
At distances $r \geq r_{{\rm Al}}$ the Al charge is $\tilde{q}_{\rm
Al}=1.8$, thereby producing charge neutrality.  For the cut--off radius
$r_{\rm Al}$, the value $r_{\rm Al}=6$\,\AA~is chosen.  The parameter
$d_{\rm Al}$ is set to 2\,\AA$^2$ for $q_{\rm Al}(r)$ in the Al--O
interactions and to 1.47\,\AA$^2$ for $q_{\rm Al}(r)$ in the Al--Al and
the Al--Si interactions. Finally, we have chosen $C_{\rm Al}=0.0653609$
for the Al--O interactions and $C_{\rm Al}=0.0637977$ for the Al--Al
and Al--Si interactions. The parameters for $d_{\rm Al}$ and
$C_{\rm Al}$ are slightly different from those reported in 
Winkler {\it et al.}~(2004). However, these slight differences are
not relevant with respect to the resulting structural and dynamic
properties.

From Eq.~(\ref{eq1}) it becomes obvious that at small distances the
potential between the Al (or Si) and the O atoms goes to minus infinity
(since the coefficients $C_{\alpha\beta}$ are positive), i.e.~it becomes
unphysical. Therefore we have modified the potential at short distances
by substituting it by a polynomial continuation that makes the potential
repulsive at very short distances:
\begin{equation}
  \phi_{\alpha {\rm O}} (r) =
  a_{1, \alpha} + a_{2, \alpha} \, r + \frac{1}{3} a_{3, \alpha} (r- a_{4, \alpha})^3
  \quad \quad r < r_{\rm c}^{\rm poly}
  \label{eq1e}
\end{equation}
with $\alpha= {\rm Si, Al, O}$. The values for $a_{1, \alpha}$, $a_{2,
\alpha}$, $a_{3, \alpha}$, $a_{4, \alpha}$, and the cut--off radius
$r_{\rm c}^{\rm poly}$ are listed in Table \ref{table1}. One may wonder
why the polynomial terms in Eq.~(\ref{eq1e}) are appropriate to describe
the interactions at small distances accurately. But even at the highest
considered temperature, $T=6000$\,K, only a negligible number of ion pairs
approach each other at distances $r < r_{\rm c}^{\rm poly}$. Thus, the
potentials as given by Eq.~(\ref{eq1e}) are only of technical importance
and do not affect at all the physical properties in the temperature
range considered in this study.

Having described in detail the model potential, we give now the main
details of the simulation runs.  Molecular dynamics simulations were done
for the two compositions 29(Al$_2$O$_3$)197(SiO$_2$) (with about 13\,mol\%
Al$_2$O$_3$) and 65(Al$_2$O$_3$)73(SiO$_2$) (with about 47\,mol\%
Al$_2$O$_3$).  In the following, we will abbreviate these compositions
as A29S197 and A65S73, respectively. Both systems were simulated at two
different mass densities $\rho$ to study how pressure affects structural
and dynamic properties. For A29S197, the densities $\rho=2.29$\,g/cm$^3$
and $\rho=2.4$\,g/cm$^3$ were chosen, which are slightly below and
above the experimental value at room temperature, $\rho=2.36$\,g/cm$^3$
(Morikawa {\it et al.}, 1982). In the case of A65S73, the densities
$\rho=2.35$\,g/cm$^3$ and $\rho=2.65$\,g/cm$^3$ were chosen, both of
which are below the experimental value $\rho=2.74$\,g/cm$^3$ (Morikawa
{\it et al.}, 1982).  In the following, we refer to runs at low and high
density by the abbreviations LD and HD, respectively.

The simulated systems consist of $2208$ and $2176$ atoms for A29S197
and A65S73, respectively, that sit in a cubic simulation box with
periodic boundary conditions in all three Cartesian directions.
The equations of motion were integrated with the velocity form of the
Verlet algorithm, and the Coulombic contributions to the potential and
the forces were calculated via Ewald summation (Binder {\it et al.},
2004). The time step of the integration was $1.6$~fs. For each of
the four systems (i.e.~A29S197 and A65S73 at two different densities
each), 13 temperatures were considered in the interval 6000\,K\,$\ge
T >$ 2000\,K (the lowest temperatures were 2470\,K and 2390\,K for
the LD and HD runs of A29S197, respectively, and 2190\,K and 2060\,K
for the LD and HD runs of A65S73, respectively).  The temperature of
the systems was controlled by coupling them to a stochastic heat bath,
i.e.~by periodically substituting the velocities of the particles with the
ones from a Maxwell--Boltzmann distribution with the desired temperature.
This thermostat has been first proposed by Andersen (1980) who has shown
also that it generates a canonical distribution in phase space.
After the system was equilibrated at the target temperature, we continued
the run in the microcanonical ensemble, i.e.~the heat bath was switched
off. In order to improve the statistics we have done eight independent
runs at each temperature. At the lowest temperatures, equilibration runs
between 23.4\,ns and 42.3\,ns real time were done, followed by production
runs of the same length. The longest simulations were done for the HD
A29S197 system at $T=2390$\,K, where the total simulation time was
$2\times42.3$\,ns (equilibration+production)$\times 8 = 676.8$\,ns,
corresponding to about 416 million time steps.  In addition to the
simulations of fully equilibrated melts, we produced glass structures by
quenching the systems from the melt to 300\,K, followed by relaxation runs
of 10,000 time steps at that temperature.  Note that the total simulation
time of the present study was 48.2 CPU years on an IBM Regatta at the
NIC J\"ulich.

\begin{figure}[tb]
\includegraphics[width=120mm]{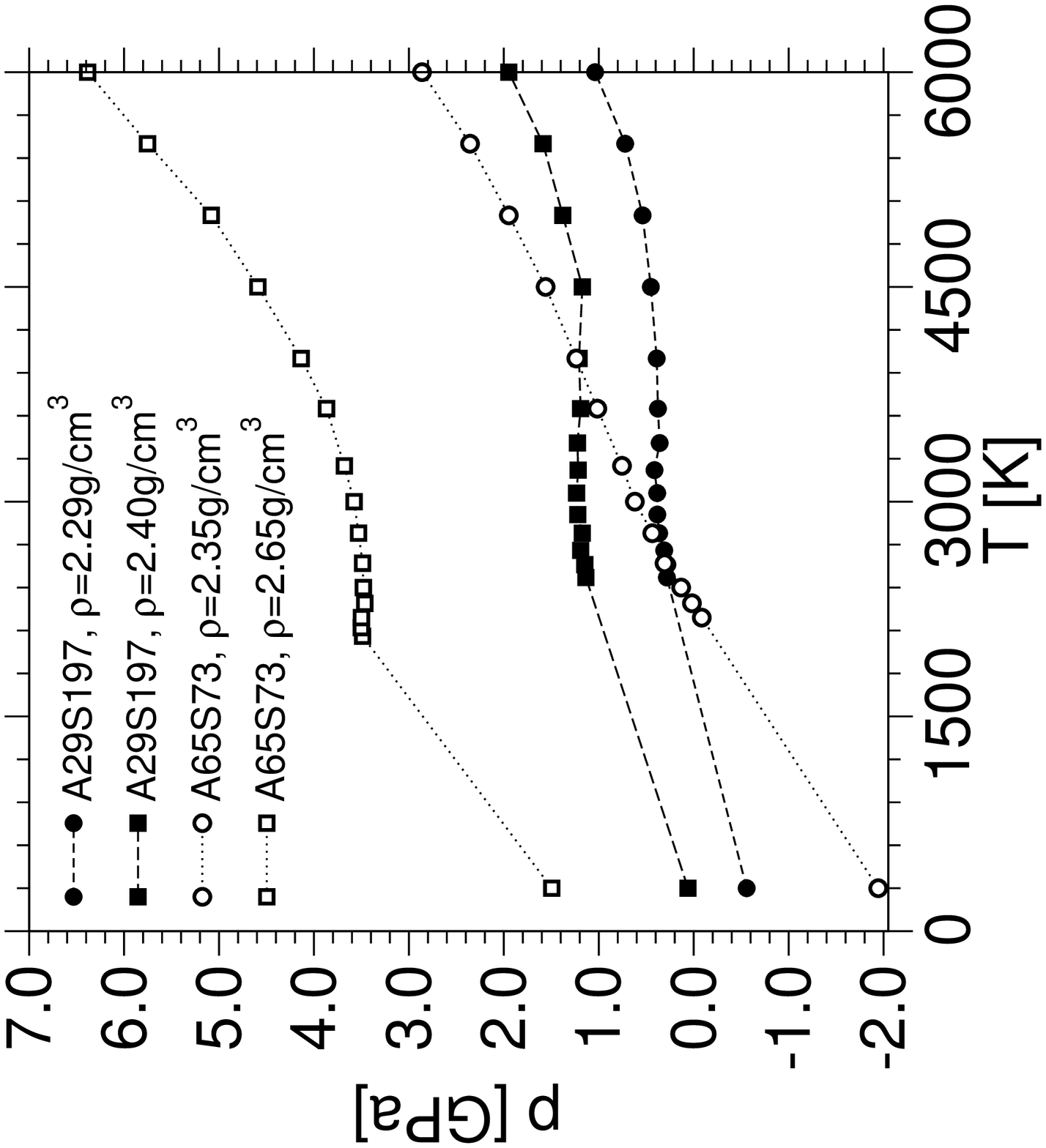}
\caption{\label{fig1}}
\end{figure}

Fig.~\ref{fig1} displays the temperature dependence of the pressure
$p$ for the different systems. In the case of the HD samples, $p(T)$
exhibits a minimum which moves to lower temperatures with increasing
Al$_2$O$_3$ concentration, namely from about 4300\,K in A29S197 to
about 2400\,K in A65S73. This feature is less pronounced in the LD
samples, and it is even absent in the LD sample of A65S73. In agreement
with density measurements of aluminium silicate melts around 2000\,K,
the pressure of the A29S197 model exhibits a relatively weak dependence
on temperature. From Fig.~\ref{fig1}, one can infer that around 2000\,K,
the density at ambient pressure is about 2.3\,g/cm$^3$. This value is
in good agreement with the experimental value of 2.32\,g/cm$^3$ at
2000\,K for an aluminium silicate melt with 14.82\,mol\% Al$_2$O$_3$
(Aksay {\it et al.}, 1979). However, for the Al$_2$O$_3$--SiO$_2$ melt
with 47\,mol\% Al$_2$O$_3$ (which is similar to our A65S73 model), Aksay
{\it et al.}~(1979) obtained a value of 2.62\,g/cm$^3$ at 2200\,K. At
this temperature, one can estimate an ambient pressure density around
2.45\,g/cm$^3$ for our simulation model and thus, in the case of the
A65S73 model, the density is about 10\% smaller than in the corresponding
real system.

\section{Results}
\label{sec3}
\subsection{Structural properties}

In this section, structural properties of the models of A29S197 and
A65S73 are investigated. The central issue is to elucidate the interplay
between the local chemical ordering around the Al atoms and the IRO that
we have recently encountered in a simulation study of AS2 (Winkler {\it
et al.}, 2004).

As we have already mentioned in the Introduction, the local structure
of amorphous aluminium silicates has been investigated by various
experimental techniques.  In particular, NMR experiments yield
detailed information about the Al--O coordination and the occurrence of
triclusters. All these experiments have been done at room temperature,
i.e.~well below the experimental glass transition temperature $T_{\rm
g}$. Therefore, a direct comparison of, e.g., NMR results with those
from a computer simulation is not very meaningful, since the glass
structures from a MD simulation have a very different thermal history
than the experimental ones (which is due to the much shorter time scale
that is accessible in a simulation).  All one can do is to extrapolate
the properties of fully--equilibrated samples at high temperatures (in
this study above 2000\,K) to temperatures at which the system would fall
out of equilibrium on the typical experimental time scale.

Keeping this issue in mind, we discuss now the temperature dependence of
the Al--O coordination in the A29S197 and A65S73 models for different
coordination numbers $z_{\rm Al-O}$.  The coordination number $z_{\rm
Al-O}$ is defined as the number of O atoms surrounding an aluminium atom
within a distance $r \le r_{\rm cut}$. For $r_{\rm cut}$, we have chosen
2.32\,\AA, which corresponds to the location of the first minimum in the
partial pair correlation function of the Al--O correlations. Note that
$r_{\rm cut}$ is significantly larger than the mean distance between
an Al and an O atom which is $r_{\rm Al-O}=1.67$\,\AA\ for our model
system. But this difference between $r_{\rm cut}$ amd $r_{\rm Al-O}$
is normal for a melt structure where the relatively large width of the
first peaks in the pair correlation function is due to structural disorder
and the thermal motion of the atoms.

Fig.~\ref{fig2} shows the coordination number distribution
$P_{\rm Al-O}(z)$ in a semi--logarithmic plot as a function of inverse
temperature for $z_{\rm Al-O}=3, 4, 5$. We have not included the results
for $z_{\rm Al-O}=6$ since six--fold coordinated aluminium atoms are
quite rare, i.e.~not more than 2\% of them are found in each of the
considered systems.  From our data it is hard to extrapolate to lower
temperatures since none of the curves can be described by an Arrhenius law
(a straight line in Fig.~\ref{fig2}) over a large temperature range. It
might even be that, e.g., the curves for $z=4$ exhibit a maximum at large
values of $1/T$, or those for $z=5$ a minimum. At least it seems that
three--fold coordinated Al atoms disappear at low temperature whereas
one may expect a significant fraction of five--fold coordinated Al atoms
also at low temperature, in particular in the HD A65S73 system. However,
Fig.~\ref{fig2} shows that for A29S197, as well as for A65S73, most of the
Al atoms are four--fold coordinated by O atoms.

Since five-- and six--fold coordinated Al atoms are relatively rare, we
expect a large number of triclusters of the form O--3(Si,Al) in order to
yield charge neutrality in the local environment of Al atoms. And indeed,
as can be seen in Fig.~\ref{fig3}a, the number of triclusters O--3(Si,Al)
per Al atom is of the order of one in the whole temperature range that is
considered for the different systems.  The minimum
in the curves for A29S197 around 3000\,K can be easily explained by a
closer inspection of the data.  At high temperature, triclusters with only
one or no Al atom are the most frequent ones. But their number strongly
decreases with decreasing temperature, whereas the number of triclusters
with two or three Al atoms increases with decreasing temperature, thus
leading to the minima in Fig.~\ref{fig3}a.  Fig.~\ref{fig3}b displays the
number of two--fold rings (i.e.~edge--sharing tetrahedra) as a function
of temperature. Clearly, there are about 0.4 two--fold rings per Al
atom at low temperature. Most of the two--fold rings contain two 
Al atoms (about 60--80\%) or one Al and one Si atom (about 20--40\%).
The fraction of edge--sharing SiO$_4$ tetrahedra is for all the considered
systems of the order of 1\% or smaller (of course, this holds only at
low temperatures). The O atoms of the two--fold rings are most
likely also triclusters, in agreement with our previous
study (Winkler {\it et al.}, 2004 and Tossell and Horbach, 2005).

In our recent work on AS2, we have demonstrated that the different chemical
ordering of aluminium and silicon leads to a microphase separation on
intermediate length scales of the order of 10--12\,\AA~whereby an Al--O rich
network percolates through the Si--O network. These structural
correlations are reflected by a prepeak in partial static structure
factors at a wavenumber around $q=0.5$\,\AA$^{-1}$. Appropriate quantities
to study the latter feature are the static concentration--concentration
structure factors $S_{c_{\alpha}c_{\alpha}}(q)$. In the following, we
define these quantities and we discuss their behavior in the case of the
aluminium silicates considered here.

For the definition of the $S_{c_{\alpha}c_{\alpha}}(q)$ we follow a
paper by Bl\'etry (1976) that generalizes the Bhatia--Thornton formalism
for two--component systems to arbitrarily many components.  Consider a
mixture of $n$ chemical species that contains a total number of $N =
\sum_{\alpha=1}^n N_{\alpha}$ particles, where $N_{\alpha}$ denotes the
number of particles of species $\alpha$.  The local number density in
reciprocal space for particles of type $\alpha$ is defined as follows
(Hansen, 1986):
\begin{equation}
  \rho_{\alpha}({\bf q}) =
   \sum_{k=1}^{N_\alpha} \exp( i {\bf q} \cdot {\bf r}_k )
  \label{eqscc_1}
\end{equation}
with ${\bf q}$ the wavevector and ${\bf r}_k$ the position of
the $k$'th particle of type $\alpha$. The partial static structure
factors are then given by (Hansen, 1986)
\begin{equation}
  S_{\alpha \beta}(q) = \frac{1}{N} \left< \rho_{\alpha}({\bf q})
  \rho_{\beta}(-{\bf q}) \right> \  .  \label{eqscc_2}
\end{equation}
Here we assume that the system is isotropic and thus, each of the $S_{\alpha
\beta}(q)$ depends only on the magnitude of the wavevector ${\bf
q}$.  The functions $S_{\alpha \beta}(q)$ are obviously symmetric,
i.e.~$S_{\alpha \beta}(q)=S_{\beta \alpha}(q)$, and therefore, there are
$n(n+1)/2$ independent partial structure factors. With this definition,
Eq.~(\ref{eqscc_2}), the $S_{\alpha \beta}(q)$ approach in the limit $q \to \infty$
the mean concentration $x_{\alpha} = N_{\alpha}/N$ for $\alpha =
\beta$ and zero for $\alpha \neq \beta$.  We show below that the
concentration--concentration structure factors can be written as linear
combinations of the partial structure factors $S_{\alpha \beta}(q)$.

The local concentration variables for particles of type $\alpha$ are
given by
\begin{equation}
   c_{\alpha}({\bf q}) = \rho_{\alpha}({\bf q})
   - x_{\alpha} \sum_{\beta=1}^{n} \rho_{\beta}({\bf q}) \ .
   \label{eqscc_3}
\end{equation}
The densities $c_{\alpha}({\bf q})$ express the local deviation from
a homogeneous density distribution of particles of type $\alpha$ and
thus, if $\rho_{\alpha}$ is equal to $N_{\alpha}/N$ for $\alpha=1,...,
n$, the variables $c_{\alpha}$ vanish.  The partial structure factors
that correspond to the concentration densities $c_{\alpha}({\bf q})$
are defined in a similar way as the partial structure factors for the
number densities:
\begin{equation}
   S_{c_{\alpha} c_{\beta}} (q) =
   \frac{1}{N} \left< c_{\alpha} ({\bf q}) c_{\beta}(-{\bf q}) \right>  \  .
   \label{eqscc_4}
\end{equation}
In this case $S_{c_{\alpha} c_{\beta}} (q) = S_{c_{\beta} c_{\alpha}}
(q)$ also holds. Moreover, the functions $S_{c_{\alpha} c_{\beta}} (q)$
obey the sum rule
\begin{equation}
   \sum_{\beta = 1}^{n} S_{c_{\alpha} c_{\beta}} (q) = 0
   \label{eqscc_5}
\end{equation}
which follows directly from the definition, Eq.~(\ref{eqscc_4}).

In the case of $n=2$ one has $c_1 = - c_2$, and thus $S_{c_1 c_1} (q)
= S_{c_2 c_2} (q)$. Furthermore, Eq.~(\ref{eqscc_5}) yields $S_{cc}
\equiv S_{c_1 c_1} (q) = - S_{c_1} c_{c_2} (q)$.  This means that for
$n=2$ there is only one relevant structure factor $S_{cc}(q)$ for the
concentration density correlations, and this quantity can be written as a linear
combination of the partial structure factors as given by Eq.~(\ref{eqscc_2}),
\begin{equation}
   S_{cc}(q) = x_2^2 S_{11}(q) + x_1^2 S_{22}(q)
               - 2 x_1 x_2 S_{12}(q) \  .
   \label{eqscc_6}
\end{equation}
For $q \to \infty$ this function approaches $x_1 x_2$, which is expected for
an ideal mixture.

Slightly more complicated is the case $n=3$.  Now the functions
$S_{c_{\alpha} c_{\beta}}(q)$ with $\alpha = \beta$ are related to the
partial number density structure factors as follows:
\begin{eqnarray}
   S_{c_1 c_1} (q) & = & (x_2 + x_3)^2 S_{11}(q)
             - 2 x_1 (x_2 + x_3) \left[ S_{12}(q) + S_{13}(q) \right] \nonumber \\
        & &     + x_1^2 \left[ S_{22}(q) + 2 S_{23}(q) + S_{33}(q) \right] \ , 
       \label{eqscc_7} \\
   S_{c_2 c_2} (q) & = & (x_1 + x_3)^2 S_{22}(q)
             - 2 x_2 (x_1 + x_3) \left[ S_{12}(q) + S_{23}(q) \right] \nonumber \\
        & &     + x_2^2 \left[ S_{11}(q) + 2 S_{13}(q) + S_{33}(q) \right] \ , 
      \label{eqscc_8} \\
   S_{c_3 c_3} (q) & = & (x_1 + x_2)^2 S_{33}(q)
             - 2 x_3 (x_1 + x_2) \left[ S_{13}(q) + S_{23}(q) \right] \nonumber \\
        & &     + x_3^2 \left[ S_{11}(q) + 2 S_{12}(q) + S_{22}(q) \right] 
     \label{eqscc_9} \ .
\end{eqnarray}
Each of the $S_{c_{\alpha} c_{\beta}}(q)$ with $\alpha \neq \beta$
can be written as a linear combination of the three $S_{c_{\alpha}
c_{\alpha}}(q)$,
\begin{eqnarray}
  S_{c_1 c_2}(q) & = & \frac{1}{2}
              \left[S_{c_3 c_3}(q) - S_{c_1 c_1}(q)
                    - S_{c_2 c_2}(q) \right] \ , \label{eqscc_10} \\
  S_{c_1 c_3}(q) & = & \frac{1}{2}
              \left[S_{c_2 c_2}(q) - S_{c_1 c_1}(q)
                    - S_{c_3 c_3}(q) \right]  \ , \label{eqscc_11} \\
  S_{c_2 c_3}(q) & = & \frac{1}{2}
              \left[S_{c_1 c_1}(q) - S_{c_2 c_2}(q)
                    - S_{c_3 c_3}(q) \right] \ . \label{eqscc_12}
\end{eqnarray}
Thus, the latter functions do not contain any additional information,
and so we consider only the correlation functions as given by
Eqs.~(\ref{eqscc_7})--(\ref{eqscc_9}) in the following.

The three functions $S_{c_{\alpha} c_{\alpha}} (q)$ are shown in
Fig.~\ref{fig4} for A29S197 and in Fig.~\ref{fig5} for A65S73.
Their behavior is qualitatively very similar in the two systems:
In both cases, the functions exhibit a pronounced peak around
$q_1=2.72$\,\AA$^{-1}$. This peak indicates the chemical ordering on
local length scales: Between nearest cation neighbors there is always
a shell of oxygen atoms, i.e.~strong fluctuations occur on length
scales of the order of the nearest cation--oxygen distance (note that
$2\pi/q_1=2.3$\,\AA~corresponds approximately to the location of the
first minima in the partial pair correlation function of Si--O and Al--O
correlations). Apart from the peak at $q_1$, in $S_{c_{\rm Al} c_{\rm
Al}} (q)$ and $S_{c_{\rm Si} c_{\rm Si}} (q)$, a pronounced prepeak
is found around $q_{\rm p}=0.5$\,\AA$^{-1}$, whereas in $S_{c_{\rm O}
c_{\rm O}} (q)$, such a prepeak is absent. This behavior of the different
concentration--concentration functions is similar to our recent result for
AS2. There, we have related the prepeak to a microsegregation into Al--O
rich and Si--O rich regions on the length scale of about 1\,nm. Indeed,
in the snapshots of Figs.~\ref{fig4} and \ref{fig5}, one can clearly
identify Al--O rich percolating regions in the Si--O network. It is
remarkable that the location of the prepeak at $q_{\rm p}$ seems to
depend neither on the Al--O concentration nor on the pressure (at least
for the pressures that are observed at the two densities considered for
each system). This finding is contrary to the behavior of structural
features on local length scales where we have found a strong dependence
on temperature, pressure, and Al$_2$O$_3$ concentration (see above).

The determination of $S_{c_{\alpha}c_{\alpha}}(q)$ from scattering
experiments is very difficult since it requires the measurement of
partial structure factors $S_{\alpha \beta}(q)$. For silicates, this
has been possible only for a few examples, e.g.~a calcium silicate glass
(Gaskell {\it et al.}, 1992). In the case of aluminium silicate glasses,
total structure factors have been measured for different compositions
using X--ray scattering by Morikawa {\it et al.}~(1982) and by
Okuno {\it et al.}~(2005). Their results can be directly compared to
simulation data. To this end, an X--ray scattering structure factor
$S_{\rm X}(q)$ has to be calculated from the partial structure factors,
$S_{\alpha \beta}(q)$, by weighting them with X--ray form factors,
\begin{equation}
   S_{\rm X}(q) = 
     \frac{N}{\sum_{\alpha} N_{\alpha} f_{\alpha}^2(s)}
     \sum_{\alpha \beta} f_{\alpha}(s) f_{\beta}(s) S_{\alpha \beta}(s)
     \label{eq_sxq}
\end{equation}
with $\alpha, \beta = \{ {\rm Si, Al, O} \}$. The form factors
$f_{\alpha}(s)$ depend on the wavenumber $q$ via $s=q/(4 \pi)$.  Note
that the $f_{\alpha}(s)$ are taken from the literature (International
Tables, 1974). Fig.~\ref{fig6} shows the ``reduced'' X--ray structure
factor $q(S_{\rm X}(q)-1)$ in comparison to experimental results. As we
recognize from the figures a good agreement is obtained between simulation
and experiment both for A29S197 and A65S73.  For the latter system,
two experimental data sets are available and, as we see in Fig.~\ref{fig6}b,
the simulation curve is closer to the more recent result of Okuno {\it
et al.}~(2005).

As we can infer from Fig.~\ref{fig6}, no prepeak is visible around
0.5\,\AA$^{-1}$ in the reduced X--ray structure factor. Note that this
is also the case for $S_{\rm X}(q)$ itself. This is due to the fact that
the main contribution to $S_{\rm X}(q)$ comes from $S_{\rm O-O}(q)$ which
does not exhibit a prepeak at 0.5\,\AA$^{-1}$. Also in neutron scattering
experiments, one has no access to partial structure factors for aluminium
silicates (due to the lack of appropriate isotopes). Thus, the accessible
total structure factor is dominated by the O--O correlations and will
hardly show a prepeak. It remains a challenge to the experimentalists
to verify the presence of the latter prepeak in aluminium silicates.

\subsection{Dynamical properties}

We have seen that the considered aluminium silicates exhibit a chemical
ordering on length scales of about 1\,nm which can be described by a
microphase separation into an Al rich and a Si rich network structure. In
our recent study of AS2 (Winkler {\it et al.}, 2004), we have shown
that the addition of Al$_2$O$_3$ to a silica melt enables a much faster
selfdiffusion of all components compared to pure SiO$_2$. This is also
the case for A29S197 and A65S73 which we consider here. In
Fig.~\ref{fig7}a, the selfdiffusion constants of our systems (for
the LD samples) are displayed on a semilogarithmic scale as a function
of inverse temperature.  Clearly, the diffusion constants increase with
increasing Al$_2$O$_3$ concentration (note that the diffusion constants of
AS2 fall between the curves shown in Fig.~\ref{fig7}a, as expected; see
Winkler {\it et al.}, 2004). In Fig.~\ref{fig7}b, the oxygen diffusion
constants $D_{\rm O}$ are shown, now also with the results for the
HD systems. We can infer from this figure that an anomalous behavior of
$D_{\rm O}$ with respect to pressure emerges: The diffusion becomes faster
with increasing pressure.  This anomaly is well--known in many different
``simple'' network forming glasses such as SiO$_2$, H$_2$O, etc.~(see,
e.g., Shell {\it et al.}, 2002 and references therein). For SiO$_2$,
this anomaly has been related to structural changes that occur with
increasing pressure (Angell {\it et al.}, 1982; Kubicki and Lasaga, 1988;
Rustad {\it et al.}, 1990; Barrat {\it et al.}, 1997): The network structure
changes gradually from a four--fold coordination of silicon atoms at
low pressure to an imperfect five--fold coordination at intermediate
pressure and eventually to a six--fold coordination at very high pressure
(note that in the latter regime the selfdiffusion constants decrease
with increasing pressure).  Intuitively, the anomalous diffusion can be
understood as follows: In the tetrahedral network structure there are
not as many possibilities for the formation of local defects as in a
higher--coordinated network. Since these defects facilitate diffusion
in a network structure (see, e.g., Horbach and Kob, 1999), the particles
in a higher--coordinated network tend to exhibit a faster diffusion than
in a tetrahedral network.

But what are the defects that facilitate the diffusion in the
Al$_2$O$_3$--SiO$_2$ networks? Important structural units might be
triclusters and five--fold coordinated silicon atoms, both of which are
more frequent in the HD systems (see above). Thus, they might be
responsible for the higher diffusion constants, as compared to the LD systems.

As we see in Fig.~\ref{fig7}a, silicon is the slowest component and
aluminium the fastest component of the diffusing species. In order to
quantify the temperature dependence of the different diffusion constants,
we show in Fig.~\ref{fig8} the ratios $D_{\rm Si}/D_{\rm O}$ and $D_{\rm
Al}/D_{\rm O}$ as a function of temperature for the different systems. It
is remarkable that $D_{\rm Si}/D_{\rm O}$ depends only weakly on density
and composition (note that also $D_{\rm Si}/D_{\rm O}$ for BKS--SiO$_2$
would fall roughly on top of the corresponding curves in Fig.~\ref{fig8},
see Winkler {\it et al.}, 2004). This shows that the diffusion of silicon
and oxygen is intimately connected with each other. On the other hand,
$D_{\rm Al}/D_{\rm O}$ nearly approaches the constant value of one in
A65S73 in the considered temperature range (at least for the HD system).
This might be due to the fact that in the Al$_2$O$_3$ rich system, A65S73,
oxygen diffusion steps occur mainly in the vicinity of Al rich regions,
in particular near triclusters, whereas in A29S197, also the O diffusion
in the slow Si rich regions is important. This issue has to be clarified
in future studies.

\section{Summary} 

Large scale molecular dynamics computer simulations were used to study
the structure and diffusion dynamics of the aluminium silicates A29S197
and A65S73. The microscopic interactions between the ions were described
by a simple pair potential proposed by Kramer {\it et al.} (1991). In
this work, we have demonstrated that this model potential yields good
agreement with available experimental data such as the mass density
and the X--ray structure factor.  However, recent NMR experiments
(Sen and Youngman, 2004; Stebbins {\it et al.}, 2005) observed a
significant number of six--fold coordinated Al atoms already at small
Al$_2$O$_3$ concentrations. By contrast, our simulation model
predicts a vanishing number of AlO$_6$ units in the glass structure. This might
be realistic for A29S197 but not for the Al--rich A65S73 system. In
the case of five--fold coordinated Al atoms, an extrapolation from
the melt data to the experimental glass transition temperature is not
easy (see above).  However, as a typical local structure, we
find tricluster O atoms in conjunction with two--fold rings. Recent
Hartree--Fock calculations that used small clusters from our recent MD
simulation of AS2, found evidence that the combination of triclusters
with two--fold rings is indeed realistic (Tossell and Horbach, 2005).
It should be emphasized that this finding refers only to binary
alumino silicates. The local structure may be very different
in ternary alumino silicates that contain also charge--balancing cations 
such as calcium or alkali ion species.

The different local ordering of Al and Si atoms leads to a
microsegregation on length scales of about 1\,nm. This structural
feature is reflected in a prepeak in static concentration--concentration
correlation functions (and also in partial static structure factors).
The location of the prepeak is around 0.5\,\AA$^{-1}$, independent
of Al$_2$O$_3$ concentration, pressure, and the details of the local
structure in the considered systems. This remarkable result shows that
according to our simulation model, the formation of Al rich channels
that percolate through the Si--O network is a central feature of the chemical
ordering in the system Al$_2$O$_3$--SiO$_2$. This is a prediction of our
simulation that could be tested by scattering experiments.  Prepeaks in
other silicates, that have a similar origin as the one found in our
aluminium silicate models, have been successfully identified in recent
neutron scattering experiments (see Gaskell {\it et al.}, 1991; Meyer
{\it et al.}, 2002; Meyer {\it et al.}, 2004; Kargl and Meyer, 2005).

Our analysis of the diffusion dynamics suggests that triclusters are
important for the mass transport in aluminium silicates.  Moreover, we
have found subtle differences in the dependence of the three diffusion
constants on temperature (e.g.~the ratio $D_{\rm Al}/D_{\rm O}$ is
essentially one for the HD A65S73 system, while it increases in the
case of the A29S197 systems). It would be interesting to see whether
these features can be understood in the framework of mode coupling theory.
Work in this direction is in progress.

Acknowledgments:
We thank W. Kob, F. Kargl, A. Meyer, and J. Stebbins for useful
discussions on this work. J. H. was supported by the Emmy Noether
program of the German Science Foundation (DFG), grants HO 2231/2-1 and
HO 2231/2-2. Computing time on the JUMP at the NIC J\"ulich is gratefully
acknowledged.

\section{References}

\begin{trivlist}
\item[]
Aksay, I. A., Pask, J. A., and Davis, R. F., 1979.
Densities of SiO$_2$--Al$_2$O$_3$ Melts.
J. Am. Cer. Soc. 62, 332--336.
\item[]
Allen, M. P., and Tildesley, D. J., 1987.
Computer Simulation of Liquids.
Clarendon Press, Oxford. 
\item[]
Andersen, H. C., 1980.
Molecular dynamics at constant pressure and/or temperature.
J. Chem. Phys. 72, 2384--2393.
\item[]
Angell, C. A., Cheeseman, P. A., and Tamaddon, S., 1982.
Pressure Enhancement of Ion Mobilities in Liquid Silicates from Computer 
Simulation Studies to 800\,kbar.
Science 218, 885--887 (1982).
\item[]
Barrat, J. L., Badro, J., and Gillet, P., 1997.
A strong to fragile transition in a model of liquid silica.
Mol. Sim. 20, 17--25.
\item[]
Binder, K., Horbach, J., Kob, W., Paul, W., and Varnik, F., 2004.
Molecular dynamics simulations.
J. Phys.: Condens. Matter 16, S429--S453.
%

%
%\item[]
%Benoit, M., Profeta, M., Mauri, F., Pickard, C. J., and Tuckerman, M. E., 2005.
%First--principles calculation of the O--17 NMR parameters of a calcium 
%aluminosilicate glass.
%J. Phys. Chem. B 109, 6052--6060.
%

%
\item[]
Cormier, L., Calas, G., and Gaskell, P. H., 2001.
Cationic environment in silicate glasses studied by neutron diffraction
with isotopic substitution.
Chem. Geol. 174, 349--363.
\item[]
Gaskell, P. H., Eckersley, M. C., Barnes, A. C., and Chieux, P., 1991.
Medium--range order in the cation distribution of a calcium silicate glass.
Nature 350, 675--678.
\item[]
Hansen, J.--P., and McDonald, I. R., 1986.
Theory of Simple Liquids.
Academic, London.

\item[]
Horbach, J., and  Kob, W., 1999.
Static and dynamic properties of a viscous silica melt.
Phys. Rev. B 60, 3169--3181.
\item[]
Horbach, J., Kob, W., and Binder, K., 2002.
Dynamics of sodium in sodium disilicate: Channel relaxation and sodium diffusion.
Phys. Rev. Lett. 88, 125502.

\item[]
International Tables for X--ray Crystallography, edited by International
Union of Crystallography (Kynoch, 1974), Vol. 4, Chap. 2.2, pp. 71--99.
\item[]
Ispas, S., Benoit, M., Jund, P., and Jullien, R., 2002.
Structural properties of glassy and liquid sodium tetrasilicate: comparison
between ab initio and classical molecular dynamics simulation.
J. Non--Cryst. Solids 307, 946--955.
\item[]
Kargl, F., and Meyer, A., 2004.
Inelastic neutron scattering on sodium aluminosilicate melts: sodium diffusion 
and intermediate range order.
Chem. Geol. 213, 165--172.

\item[]
Kramer, G. J., de Man, A. J. M., and van Santen, R. A., 1991. 
Zeolites versus Aluminosilicate Clusters: The Validity of a Local
Description. J. Am. Chem. Soc. 64, 6435--6441.
\item[]
Kubicki, J. D., and Lasaga, A. C., 1988.
Molecular dynamics simulations of silica melt and glass: Ionic and covalent models.
Am. Min. 73, 941--955.
\item[]
Kubicki, J. D., and Toplis, M. J., 2002.
Molecular orbital calculations on aluminosilicate tricluster 
molecules: Implications for the structure of aluminosilicate glasses.
Am. Min. 87, 668--678.

\item[]
Lee, S. K., and Stebbins, J. F., 2000.
The structure of aluminosilicate glasses: High--resolution O--17 and 
Al--27 MAS and 3QMAS NMR Study.
J. Phys. Chem. B 104, 4091--4100.
\item[]
MacDowell, J. F., and Beall, G. H., 1969.
Immiscibility and Crystallization in Al$_2$O$_3$--SiO$_2$ Glasses.
J. Am. Cer. Soc. 52, 17--25.
\item[]
Meyer, A., Horbach, J., Kob, W., Kargl, F., and Schober, H., 2004.
Channel formation and intermediate range order in sodium silicate 
melts and glasses.
Phys. Rev. Lett. 93, 027801.
\item[]
Meyer, A., Schober, H., Dingwell, D. B., 2002.
Structure, structural relaxation and ion diffusion in sodium disilicate
melts.
Europhys. Lett. 59, 708--713.
\item[]
Morikawa, H., Miwa, S., Miyake, M., Marumo, F., and Sata, T., 1982.
Structural analysis of SiO$_2$--Al$_2$O$_3$ glasses.
J. Am. Cer. Soc. 65, 78--81.
\item[]
Okuno, M., Zotov, N., Schm\"ucker, M., and Schneider, H., 2005.
Structure of SiO$_2$--Al$_2$O$_3$ glasses: Combined X--ray diffraction, IR
and Raman studies. 
J. Non--Cryst. Solids 351, 1032--1038.
\item[]
Rustad, J. R., Yuen, D. A., and Spera, F. J., 1990.
Molecular dynamics of liquid SiO$_2$ under high pressure.
Phys. Rev. A 42, 2081--2089.
\item[]
Schm\"ucker, M., Schneider, H., MacKenzie, K. J. D., and Okuno, M., 1999.
Comparative $^{27}$Al NMR and LAXS Studies on Rapidly Quenched Aluminosilicate
Glasses.
J. Eur. Cer. Soc. 19, 99--103.
\item[]
Sen, S., and Youngman, R. E., 2004.
High--Resolution Multinuclear NMR Structural Study of Binary Aluminosilicate
and Other Related Glasses.
J. Phys. Chem. B 108, 7557--7564.
\item[]
Shell, M. S., Debenedetti, P. G., and Panagiotopoulos, A. Z., 2002.
Molecular structural order and anomalies in liquid silica.
Phys. Rev. E 66, 011202.
\item[]
Stebbins, J. F., Du, L.--S., and Pratesi, G., 2005.
Aluminium coordination in natural silica glasses from the Libyan
Desert (Egypt): high field NMR results.
Phys. Chem. Glasses, in press.
\item[]
Stebbins, J. F., and Xu, Z., 1997. 
NMR evidence for excess non-bridging oxygen in an aluminosilicate glass.
Nature 390, 60--62.
\item[]
Tossell, J. A., and Cohen, R. E., 2001.
Calculation of the electric field gradients at 'tricluster'-like O atoms in 
the polymorphs of Al$_2$SiO$_5$ and in aluminosilicate molecules: models for 
tricluster O atoms in glasses.
J. Non--Cryst. Solids 286, 187--199.
\item[]
Tossell, J. A., and Horbach, J., 2005.
O Tricluster Revisited: Classical MD and Quantum Cluster Results for Glasses
of Composition (Al$_2$O$_3$)2(SiO$_2$).
J. Phys. Chem. B. 109, 1794--1797.
\item[]
van Beest, B. W. H., Kramer, G. J., and van Santen, R. A., 1990.
Force Fields for Silicas and Aluminophosphates Based on {\it Ab Initio}
Calculations. Phys. Rev. Lett. 64, 1955--1958.
\item[]
Winkler, A., Horbach, J., Kob, W., and Binder, K., 2004.
Structure and diffusion in amorphous aluminum silicate: A molecular dynamics 
computer simulation.
J. Chem. Phys. 120, 384--393.
\item[]
Xue, X.Y., and Kanzaki, M., 1999.
NMR characteristics of possible oxygen sites in aluminosilicate glasses and 
melts: An ab initio study.
J. Phys. Chem. B 103, 10816--10830.
\end{trivlist}

\newpage
\section{List of Captions}

Fig.~\ref{fig1}:
Pressure as a function of temperature at the indicated compositions
and densities. Note that the points at 300\,K are far below the glass
transition temperature of the simulations, which is around 2000\,K
(see text).

Fig.~\ref{fig2}:
Temperature dependence of $P_{\rm Al-O}(z)$ for $z=3, 4, 5$ as indicated,
for a) A29S197 and b) A65S73.

Fig.~\ref{fig3}:
This plot shows the amount of triclusters, O--3(Si,Al), and 2--fold rings
as a function of temperature: a) Number of triclusters and b) number of
2--fold rings per Al atom for the different systems as indicated.

Fig.~\ref{fig4}:
Structure factors $S_{c_{\alpha}c_{\alpha}}(q)$ for A29S197 at the
temperature $T=2470$\,K and the different densities as indicated, a)
$S_{c_{\rm Si}c_{\rm Si}}(q)$, b) $S_{c_{\rm O}c_{\rm O}}(q)$, and c)
$S_{c_{\rm Al}c_{\rm Al}}(q)$ [see Eqs.~(\ref{eqscc_7}--(\ref{eqscc_9})
for the definition of $S_{c_{\alpha}c_{\alpha}}(q)$]. Also shown is a
snapshot at $T=300$\,K which illustrates the intermediate range order
as reflected in the prepeak around $q=0.5$\,\AA$^{-1}$. The large white
spheres are the silicon atoms, the large black spheres are the aluminium
atoms, and the small black spheres are the oxygen atoms. Note that the
size of the spheres does not correspond to the actual size of the atoms.

Fig.~\ref{fig5}:
The same as in Fig.~\ref{fig4} but now for A65S73.  The temperatures
are $T=2190$\,K and $T=2060$\,K for the low and the high density,
respectively.

Fig.~\ref{fig6}:
Reduced X--ray scattering structure factor $q(S_{\rm X}(q)-1)$ as
calculated from the simulation using Eq.~(\ref{eq_sxq}) in comparison
to experimental results by Morikawa {\it et al.}~(1982) and Okuno {\it
et al.}~(2005) (dashed lines), for a) A29S197 and b) A65S73.

Fig.~\ref{fig7}:
Arrhenius plots of the selfdiffusion constants $D_{\alpha}$, a) for Si,
Al, and O for the different compositions at low densities and b) for O
for all considered systems as indicated.

Fig.~\ref{fig8}:
Temperature dependence of the ratios of the diffusion constants,
$D_{\alpha}/D_{\beta}$ ($\alpha, \beta = {\rm Si, Al, O}$), for the
different systems as indicated.

\newpage
\section{List of Tables}
\begin{table}[h]
\begin{center} \begin{tabular}{|c|c|c|c|c|c|}  \hline $\alpha-{\rm O}$ & $a_{1, \alpha}$ [eV]
& $a_{2, \alpha}$ [eV/\AA] & $a_{3, \alpha}$ [eV/\AA$^{3}$] & $a_{4, \alpha}$ [\AA] &
$r_{\rm c}^{\rm poly}$ [\AA] \\ \hline \hline
  Si--O & -23.96027 & -2.85441 & -50.0 & 1.41590  & 1.276 \\ \hline
  Al--O & -87.62405 & 26.72474 & -3.0  & 4.49012  & 1.28  \\ \hline
  O--O  & -30590.38 & 90.38499 & -0.1  & 97.25877 & 1.9547 \\ \hline
\end{tabular}
\end{center}
\caption{\label{table1} Parameters for the polynomial continuation at small $r$ that prevents
         the potentials from going to minus infinity for $r \to 0$.}
\end{table}

\newpage
\section{List of Figures}

\newpage

\begin{figure}[t]
\vspace*{0.2cm}
\includegraphics[width=75mm]{fig2a}
\hspace*{0.5cm}
\includegraphics[width=75mm]{fig2b}
\caption{\label{fig2}}
\end{figure}
\begin{figure}[b]
\vspace*{0.2cm}
\includegraphics[width=75mm]{fig3a}
\hspace*{0.5cm}
\includegraphics[width=75mm]{fig3b}
\caption{\label{fig3}}
\end{figure}
\begin{figure}[tb]
\includegraphics[width=70mm]{fig4a}
\hspace*{0.5cm}
\includegraphics[width=85mm]{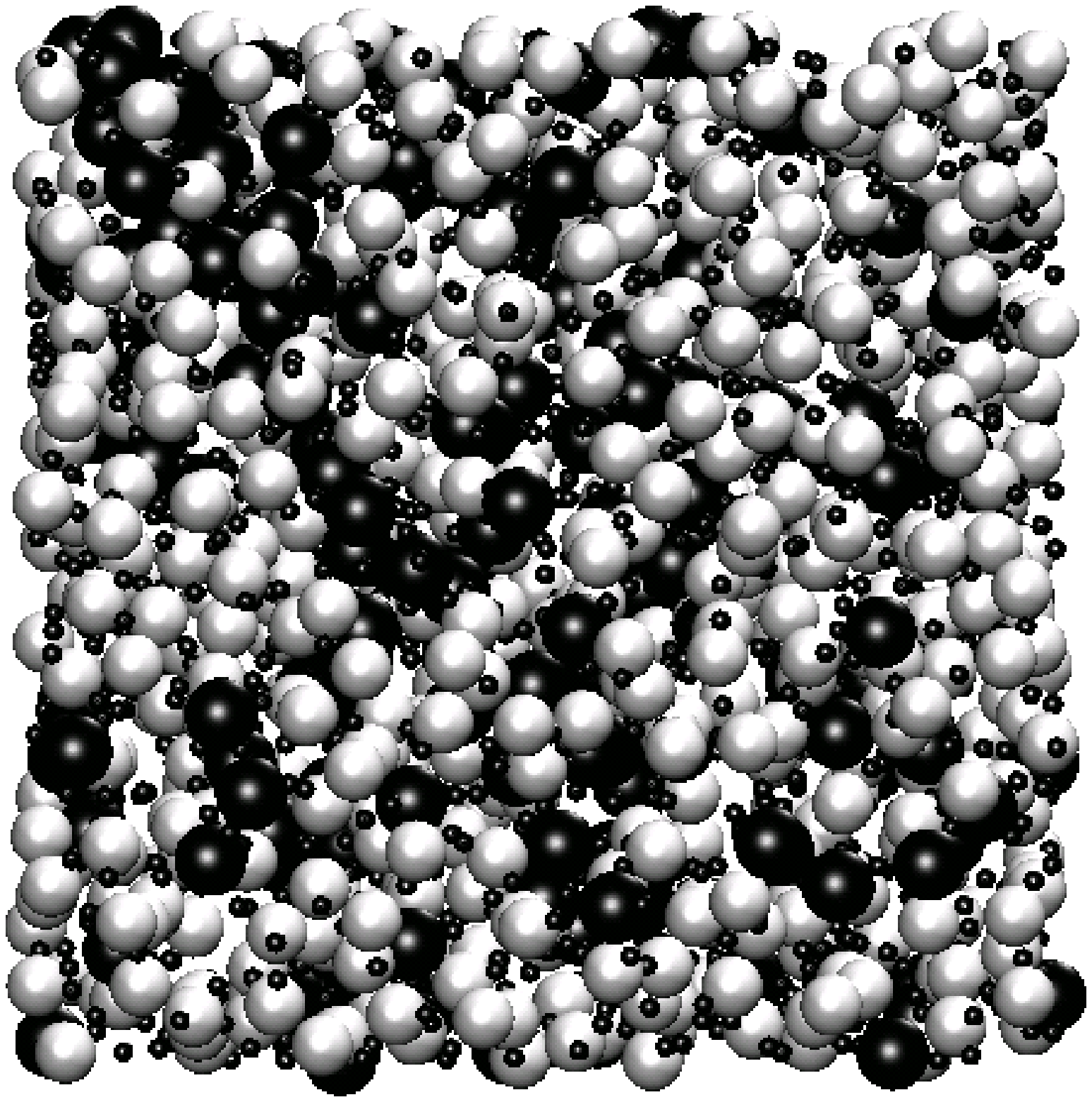}
\caption{\label{fig4}}
\end{figure}
\begin{figure}[tb]
\includegraphics[width=70mm]{fig5a}
\hspace*{0.5cm}
\includegraphics[width=85mm]{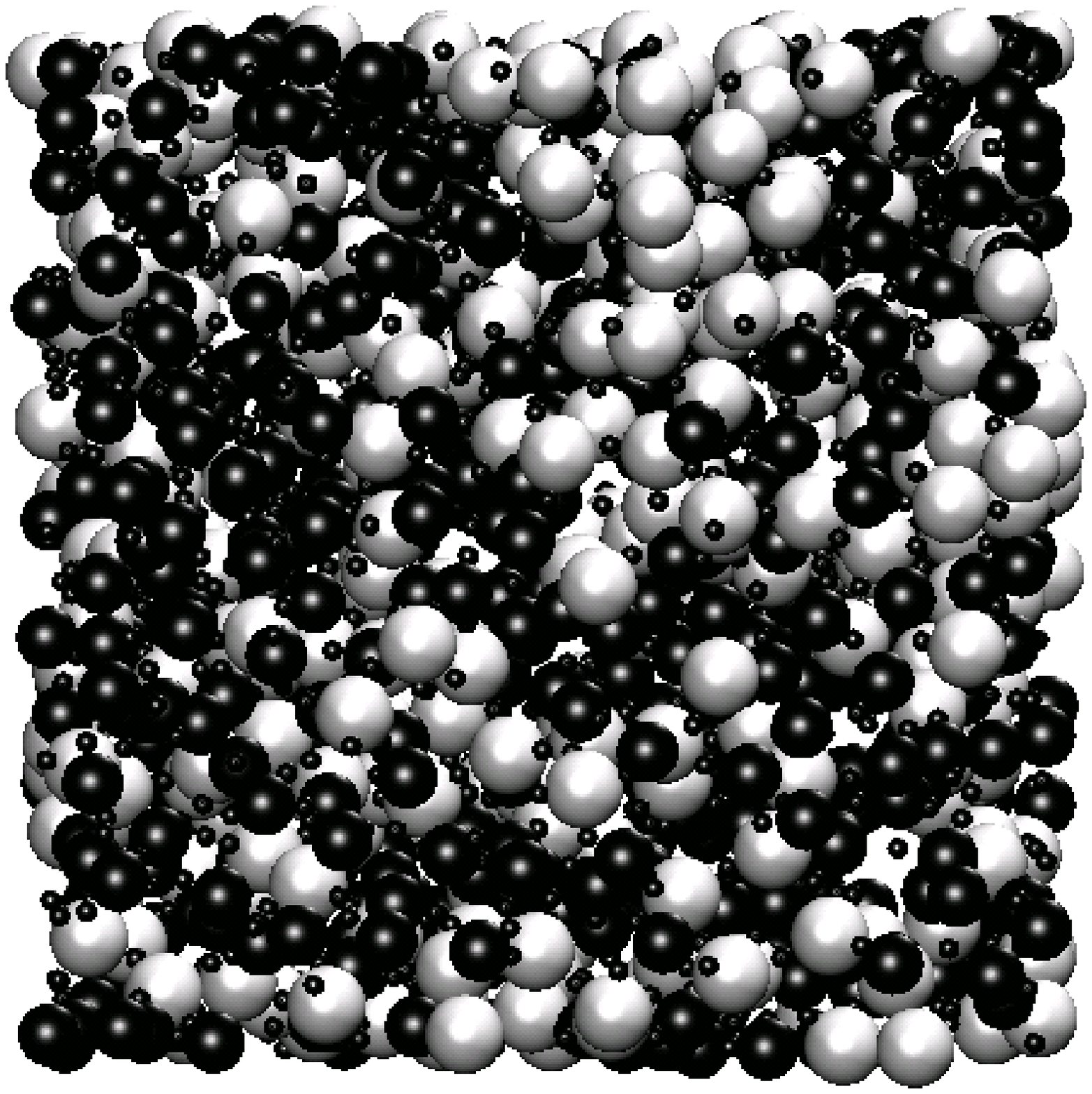}
\caption{\label{fig5}}
\end{figure}
\begin{figure}[tb]
\includegraphics[width=75mm]{fig6a}
\includegraphics[width=75mm]{fig6b}
\caption{\label{fig6}}
\end{figure}
\begin{figure}[tb]
\includegraphics[width=75mm]{fig7a}
\hspace*{0.5cm}
\includegraphics[width=75mm]{fig7b}
\caption{\label{fig7}}
\end{figure}
\begin{figure}[tb]
\vspace*{0.5cm}
\includegraphics[width=120mm]{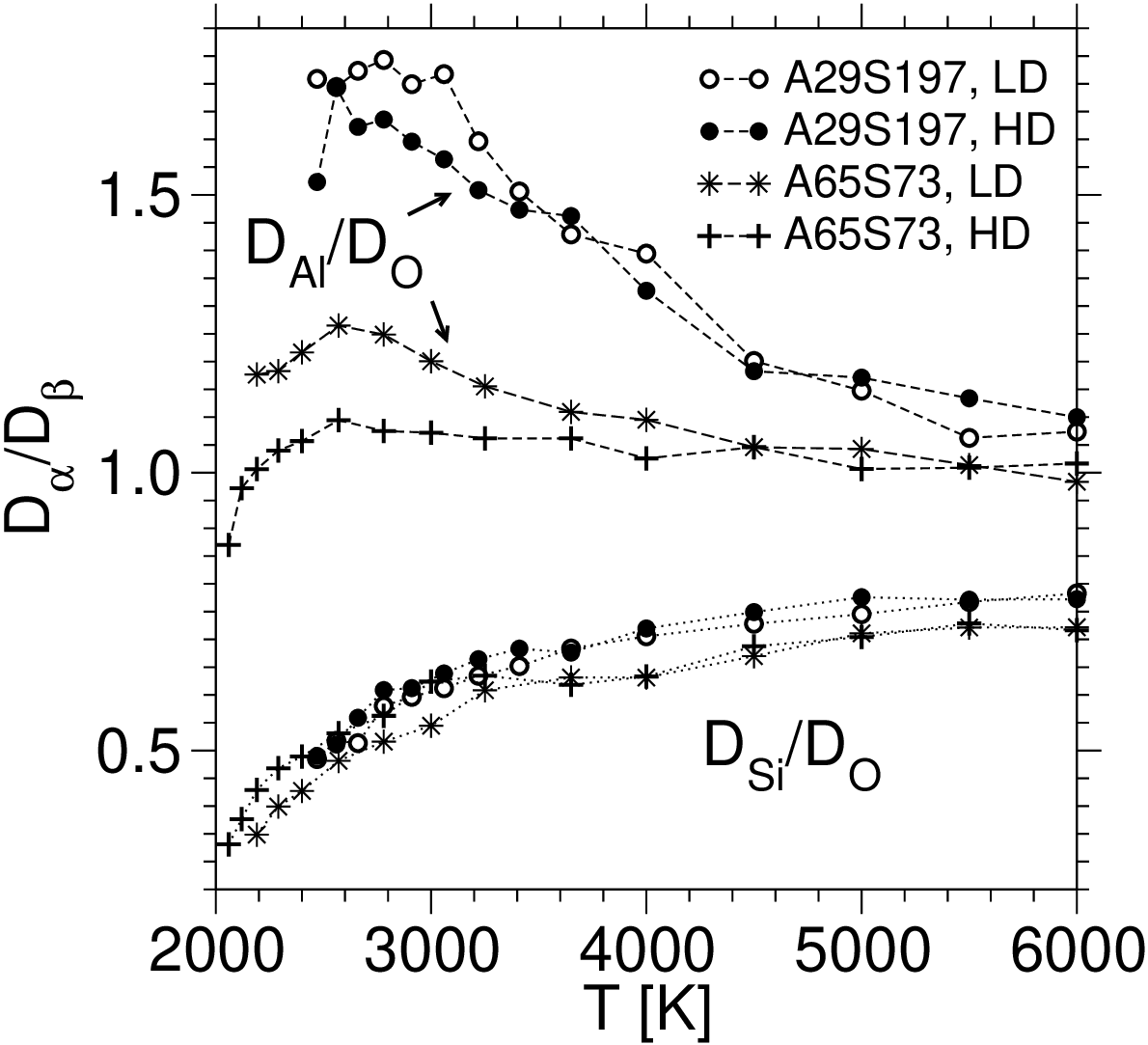}
\caption{\label{fig8}}
\end{figure}

\end{document}